\documentclass[letter]{aa} 

\usepackage{graphicx, comment}
\usepackage{commath}
\usepackage{txfonts}
\usepackage{subfig}
\usepackage{upgreek}
\usepackage{threeparttable} 
\usepackage{booktabs}       
\usepackage[colorlinks=true]{hyperref}
\hypersetup{
     colorlinks   = true,
     citecolor    = blue
}

\usepackage{orcidlink}

\newcommand{\C}{3C\,84}
\newcommand{\rotang}{-4.5^\circ}
\newcommand{\protang}{4.5^\circ}
\newcommand{\beam}{(0.17\times0.40)\,\textrm{mas}}

\newcommand{\Mj}{5.0\pm1.7}
\newcommand{\Mjrel}{6.7\pm2.0}
\newcommand{\Mx}{9.8\pm3.2}
\newcommand{\enthalpy}{0.26\pm0.20}
\newcommand{\ajj}{0.14\pm0.06}
\newcommand{\ax}{0.07\pm0.02}
\newcommand{\bw}{0.27\pm0.04}

\newcommand{\rj}{0.36\pm0.09\,\textrm{pc}}
\newcommand{\phij}{4.5^\circ\pm1.2^\circ}
\newcommand{\thetaj}{35^\circ\pm5^\circ}
\newcommand{\gj}{1.35\pm0.14}

\newcommand{\tp}{30\pm10\,\textrm{years}}
\newcommand{\chiamp}{\sim1}
\newcommand{\chialt}{\sim2}

\begin{document} 
\renewcommand{\arraystretch}{1.5}

    \title{Unravelling the dynamics of cosmic vortices: \\Probing a Kelvin-Helmholtz instability in the jet of \C}

   \author{
   G.~F. Paraschos\inst{1} \orcidlink{0000-0001-6757-3098},
   V. Mpisketzis\inst{2,3}\orcidlink{0009-0009-9682-3119} 
          }

   \authorrunning{G.~F. Paraschos et al.}
   \institute{
              $^{1}$Max-Planck-Institut f\"ur Radioastronomie, Auf dem H\"ugel 69, D-53121 Bonn, Germany\\ 
              $^{}$\ \email{gfparaschos@mpifr-bonn.mpg.de}\\
              $^2$Department of Physics, National and Kapodistrian University of Athens, Panepistimiopolis, 15783 Zografos, Greece\\
              $^3$Institut für Theoretische Physik, Goethe Universität Frankfurt, Max-von-Laue-Str.1, 60438 Frankfurt am Main, Germany\\
             }

   \date{Received -; accepted -}

 \abstract
   {
   Understanding the creation of relativistic jets originating from active galactic nuclei, require a thorough understanding of the accompanying plasma instabilities.
   Our high sensitivity, high resolution, global very long baseline interferometry observations of the jet in the radio galaxy \C\ enable us to study its inner morphology, which resembles a thread-like pattern.
   We find that this pattern can be described by a Kelvin-Helmholtz instability, consisting of four instability modes.
   Our model favours a jet described by a Mach number of $M_\textrm{j} = \Mj$ and a sound speed of $\alpha_\textrm{j} = \ajj$.
   With it, we are able to describe the internal structure of \C\ and to tentatively connect the origin of the instability to accretion disc activity. 
   }


   \maketitle

\section{Introduction}

With the increasing sensitivity of modern very-long-baseline interferometry (VLBI) arrays, more and more active galactic nuclei (AGN) jets have been identified to showcase filamentary structures.
Such structures have been mostly identified in parsec-scale jets \citep[e.g.][]{Lobanov01, Hardee03, Perucho12, VegaGarcia19, Fuentes23, Traianou24, Britzen24}; one exception is the jet of M\,87 where sub-parsec-scale structures have been identified \citep[e.g.][]{Lobanov03, Hardee11, Pasetto21, Nikonov23}.
A persistent characteristic of these sources is that they exhibit edge brightening (i.e., the outer sheath of the jet is brighter than the inner spine).

The radio galaxy NGC\,1275, harbouring the radio source \C, is another example, where a double rail structure has be identified \citep[e.g.][]{Nagai16, Giovannini18, Paraschos22, Savolainen23}.
It is one of the nearest and brightest radio sources in the northern sky and is therefore supremely suited to investigate the characteristics of the plasma that constitutes the jet.
Its bifurcated morphology, reaching deep into the core region \citep[e.g.][]{Giovannini18, Punsly21, Paraschos24}, is indicative of instabilities being present in the plasma.
Furthermore, in the sub-parsec-scale region, different velocities have been observed \citep[e.g.][]{Hodgson21, Weaver22, Paraschos22}, suggestive of Kelvin-Helmholtz (K-H) instability propagation.
\C\ also exhibits variability on the time scales of a few decades \citep[e.g.][]{Hodgson18, Paraschos23}, which might also be connected to projection effects of twisting filaments.

Such filaments provide a wealth of insights into the physics governing jets.
For example, by modelling the K-H instability modes \citep[e.g.][]{Perucho05, Mizuno07} or by testing current driven instability modes against simulations \citep[CDI; e.g.][]{Mizuno09, McKinney09} one can constrain jet propagation parameters, such as the jet's internal Mach number $M_\textrm{j}$, density ratio of the jet to the ambient medium $\eta$, sound speed of the jet and the ambient medium ($\alpha_\textrm{j}$ and $\alpha_\textrm{x}$ respectively), and the instability propagation velocity $v_\textrm{w}$.

Here we leveraged the excellent, sub-parsec-scale resolution achieved when observing \C\ with centimetre VLBI, to investigate the jet's filamentary structure in the vicinity of the central engine in great detail.
We invoke the K-H instability \citep[see, for example][]{Hardee84, Hardee86, Hardee87, Hardee97, Hardee00} to gain analytic results of the underlying modes and to make predictions about the jet morphology.

This paper is structured as follows: in Sect.~\ref{sec:Results} we briefly touch upon the data reduction and discuss our results.
In Sect.~\ref{sec:Discussion} we examine the modelling parameters of the filamentary structures of \C.
Finally, in Sect.~\ref{sec:Conclusions} we summarise our conclusions.
Throughout this work we assume a $\Lambda$ cold dark matter cosmology with $H_0 = 67.8\,\textrm{kms}^{-1}\, \textrm{Mpc}^{-1}$, $\Omega_\Lambda = 0.692$, and $\Omega_\textrm{M} = 0.308$ \citep{Planck16}.
At a $z=0.0176$ \citep[][]{Strauss92}, this is equivalent to a luminosity distance $D_\textrm{L} = 78.9 \pm 2.4\, \textrm{Mpc}$ and a parsec to milliarcsecond conversion factor $\psi = 0.36\, \textrm{pc mas}^{-1}$.

\section{Observations, methods, and results}\label{sec:Results}
\subsection{Observations}
The array configuration, data correlation and calibration, as well as imaging have already been discussed in \cite{Paraschos24b}.
In short, \C\ was observed in November 2021 at 22\,GHz by a global VLBI array consisting mainly of the European VLBI Network (EVN) in conjunction with a number of additional antennas, totalling twenty-two.
Post-processing consisted of calibrating the data with the \texttt{rPicard} \citep{Janssen19} and \texttt{polsolve} \citep{MartiVidal21} pipelines and imaging their output using the \texttt{CLEAN} algorithm within the \texttt{difmap} software package \citep{Shepherd94}.

While in \cite{Paraschos24b} the focus of the analysis was on the linearly polarised signal traced by the double thread structure mainly in the core region \citep[see also][]{Kramer24}, here we investigated these threads in the total extent of the downstream jet.
The first step to identify them was to convolve the image with a circular beam the size of 0.6\,mas, which is approximately double the size of the nominal beam $\beam$.
This step was required to blur out the finer scale structure, allowing us to robustly determine the jet ridge line\footnote{Here we defined the ridge line as the main jet axis.} \citep[see e.g.][]{Paraschos22}.
Next we fitted slices locally perpendicular to the jet axis with a bimodal function; the slices were performed on the jet image convolved with its nominal beam.
The distance between each slice (step size) was 0.1\,mas.
Our choice of a bimodal function fit was a natural consequence of the apparent double rail morphology present in \C.
Finally, we differentiated between the two peaks of the fit by assigning the higher peaked flux density ones to the green thread (`T1') and the lower peaked flux density ones to the blue thread (`T2').
The smoothed threads T1 and T2 are presented in Fig.~\ref{fig:threads}.
An alternative thread identification is discussed in Appendix~\ref{app:Alt}. 

The two threads showcase an intertwining, helical pattern suggestive of a Kelvin-Helmholtz instability.
They are most likely the result of the interplay of the plasma of the jet and the surrounding medium, spurned by an external periodic source, such as jet axis precession, accretion disc rotation, nutation, or jet wobbling \citep{Lobanov01}.
Such periodicity is also identified in relativistic magneto-hydrodynamic simulations \citep[e.g.][]{Mizuno15, Fuentes18}.
The threads can be described by the jet internal (external medium) Mach number $M_\textrm{j}$ ($M_\textrm{x}$) and velocity $\beta_j$ ($\beta_x$).
Finally, an auxiliary measurables is the ratio between the internal and external medium density $\eta_j\equiv \rho_j/\rho_x$, also called enthalpy ratio.

\begin{figure}
\centering
\includegraphics[scale=0.3]{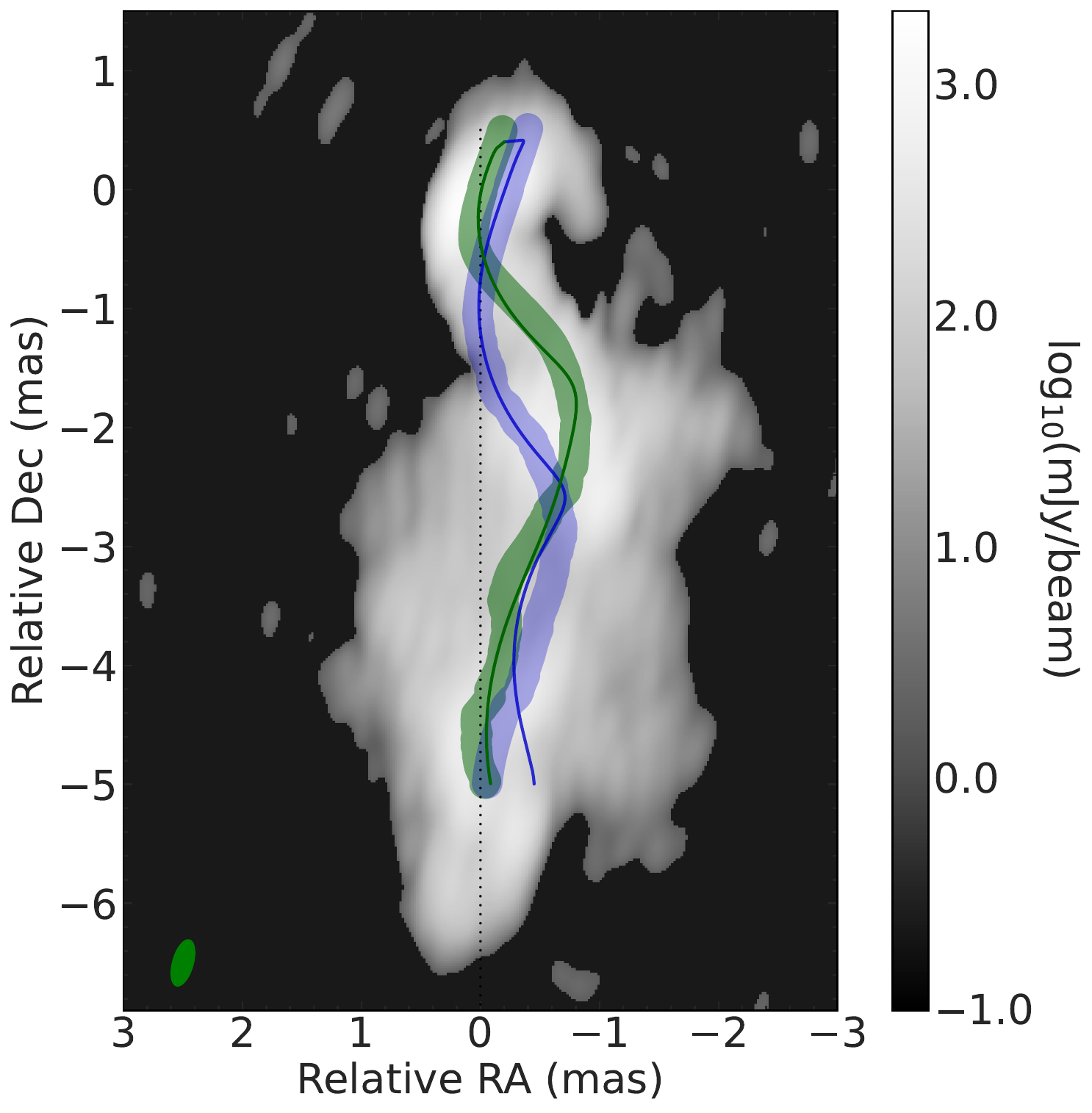}
  \caption{
    Stokes I image of \C\ showcasing a filamentary morphology.
    The colour scale is logarithmic and the units are mJy/beam.
    The dark green ellipse in the lower left corner shows the convolving beam size $\beam$.
    Superimposed are two threads; the green (blue) one traces the filament characterised by higher (lower) flux density.
    Their widths correspond to the position uncertainty.
    The computed K-H models are shown with the continuous thin lines, matching the colours of the double helix morphology.
    The dotted line corresponds to a linear fit of the ridge line with the purpose of determining the position angle ($\textrm{P. A.}=\protang$) of the jet propagation.
    The image was turned by $\rotang$ to align the jet axis with the ordinate.
    A cutoff at $\sim5$\,mas from the jet apex was implemented, to only include high $S/N$ areas in our fit.
    }
     \label{fig:threads}
\end{figure}

\begin{table}[h]
\centering
\begin{threeparttable}
\caption{Model input (top) and output (bottom) parameters}
\label{tab:inp_par}
\begin{tabular}{c|c}
\hline
\hline
Parameter & Value \\
\hline
$R_\textrm{j}$ & $\rj$\tnote{a} \\ %
$\mathcal{\phi}_\textrm{j}$ & $\phij$\tnote{$\dagger$} \\ %
$\theta_\textrm{j}$ & $\thetaj$\tnote{b} \\ %
$\Gamma_\textrm{j}$  & $\gj$\tnote{c} \\ %
$\beta_\textrm{w}$          & $\bw$\tnote{d}    \\ %
\hline
$M_\textrm{j}$              & $\Mj$      \\ %
$\mathcal{M}_\textrm{j}$    & $\Mjrel$      \\ %
$M_\textrm{x}$              & $\Mx$         \\ %
$\eta$                      & $\enthalpy$    \\ %
$\alpha_\textrm{j}$         & $\ajj$    \\ %
$\alpha_\textrm{x}$         & $\ax$  \\ %
\hline
\end{tabular}
\begin{tablenotes}
\item[$\dagger$] Calculated in this work.
\item[a] Adopted from \cite{Paraschos21}.
\item[b] Adopted from \cite{Oh22}.
\item[c] Adopted from \cite{Hodgson21}.
\item[d] Computed from \cite{Hodgson18} and \cite{Kam24}.
\end{tablenotes}
\end{threeparttable}
\end{table}

\begin{table*}[h!]
\centering
\caption{K-H modelled parameters for T1 and T2}
\begin{tabular}{c|c|c|c|c|c|c|c}
\hline
\hline
Mode & $\lambda_\textrm{i}$ & $A_\textrm{i}$ & $\phi_\textrm{i}$ & $k_\textrm{i}$ & K-H & $\epsilon_{\lambda_\textrm{i}}$ & $\lambda^*_\textrm{n,m}$ \\ 
     &  \begin{tabular}[c]{@{}c@{}} (mas) \\ \hline \end{tabular}                &   \begin{tabular}[c]{@{}c@{}} (mas) \\ \hline \end{tabular}              &   \begin{tabular}[c]{@{}c@{}} (rad) \\ \hline \end{tabular}            &   \begin{tabular}[c]{@{}c@{}} (mas) \\ \hline \end{tabular}  &   \begin{tabular}[c]{@{}c@{}} mode \\ \hline \end{tabular}  &   \begin{tabular}[c]{@{}c@{}} {} \\ \hline \end{tabular}  &   \begin{tabular}[c]{@{}c@{}} (mas) \\ \hline \end{tabular} \\
     & T1 | T2              &  T1 | T2            &  T1 | T2          &  T1 | T2 &  T1 | T2 &  T1 | T2 &  T1 | T2 \\
\hline
1  & 6.4 | 7.0  & 1.5 | 1.0 & 1.5 | 6.2 & 0.4 & $H_\mathrm{s}$  & 6\% | 17\%  & 10.0    \\ 
2  & 3.9        & 1.0 | 1.7 & 0.9 | 4.5 & 0.4 & $E_\mathrm{s}$  & 7\%         & 9.6     \\ 
3  & 2.3 | 2.7  & 0.8 | 0.8 & 5.5 | 3.7 & 0.4 & $H_\mathrm{b1}$ & 12\% | 5\%  & 8.7     \\ 
4  & 1.9        & 0.5 | 0.4 & 4.2 | 2.1 & 0.4 & $E_\mathrm{b1}$ & 8\%         & 8.6     \\ 
\hline
\end{tabular}
\caption*{\textbf{Note}: The listed parameters correspond to the wavelength, amplitude, phase, and longitudinal wavenumber as described in Eq.~\ref{eq:xa}; the K-H mode; the discrepancy between the model and measurement; and the K-H mode wavelength, respectively.}
\label{tab:fit_amp}
\end{table*}

\subsection{Methods}

In order to fit the two threads, we assumed that they are best approximated by a sum of sinusoidal modes following a helical trajectory, which we then projected on the plane of the sky.
Mathematically, this can be expressed as:
\begin{equation}
    x_a = A  \cdot \cos\left(\frac{2 \pi}{\lambda} + \varphi \right) \cdot \cos(\theta_\textrm{j}) + x \cdot \sin(\theta_\textrm{j})\label{eq:xa}
\end{equation}
and 
\begin{equation}
    y_a = A \cdot \sin\left(\frac{2 \pi}{\lambda}  + \varphi\right),
\end{equation}
where $A \propto  A_i R_\textrm{j}(z)^{k_\textrm{i}}$ is  the amplitude,  $\lambda \propto  \lambda_i R_\textrm{j}(z)^{k_\textrm{i}}$ is the wavelength and $k_\textrm{i}$ the longitudinal wavenumber, $x$ is the distance of the jet from the apex, $\varphi$ is the phase, $\theta_\textrm{j}$ is the viewing angle to our line of sight to the jet.
The amplitude ($A_\textrm{i}$) and measured wavelength ($\lambda_\textrm{i}$) are both proportional to the jet radius $R_\textrm{j}(x)/\textrm{pc}$ \citep[][]{Hardee00}.
Additionally, we assumed that jet opening half-angle $\mathcal{\phi}_\textrm{j}$ satisfies the relation $\mathcal{\phi}_\textrm{j}\ll M_\textrm{j}^{-1}$, which ensures that our computed solutions are within linear perturbation in a cylindrical jet regime \citep[e.g.][]{Hardee86, Hardee87}.

Determining the appropriate number of modes to fit required carefully considering its impact on the goodness of fit.
We chose to fit the first four modes ($\chi_\textrm{4}^2  \chiamp$) and the resulting fits are shown with the thin blue and green lines in Fig.~\ref{fig:threads}.
While the green thread fit is confined within the uncertainty band, we observe a deviation of the blue band from the observed threads, particularly in regions farther from the central black hole. 
This behaviour arises from our model selection criterion. 
Adding a fifth mode marginally decreased the formal $\chi^2$ by fitting better these low S/N regions, but it also increased the Akaike information criterion (AIC) value by a factor of $\sim1.3$, thus, indicating that an increase in model complexity was not justified.
Similarly, employing only three modes did not result in an adequate fit, indicated by a $\chi_\textrm{3}^2 = 3$.

For each mode of the fit, we then calculated the corresponding resonant K-H mode wavelength $\lambda^*_\textrm{n,m}$, given by 
\begin{equation}
    \lambda^*_\textrm{nm} = \lambda_\textrm{i} \cdot(n+2\cdot m+0.5).\label{eq:lmn}
\end{equation}
The parameters $(n,\ m)$, appearing in the above Eq.~\ref{eq:lmn} are the azimuthal wave number and its order, respectively.
For relativistic jets, the first three azimuthal numbers are expected to be most influential \citep[e.g.][]{Hardee00, Mizuno07}.
They correspond to the pinch ($n=0$), helical ($n=1$), and elliptical ($n=2$) instability, respectively.
Their order, on the other hand, controls whether the perturbation is manifested in the surface ($m=0$) or body ($m>0$) of the jet.
We note that since $\lambda^*$ depends only on the physical characteristics of the jet, its value should not differ per mode \citep[see e.g.][]{Lobanov03}.

Furthermore, when comparing the modelled wavelengths ($\lambda^*_\textrm{nm}$) to all the measured ones ($\lambda_\textrm{i}$) in order to identify each mode, we tolerated a discrepancy $\epsilon_{\lambda_\textrm{i}}$ of up to $25\%$ between  $\lambda^*_\textrm{nm}$ and $\lambda_\textrm{i}$ per mode.
This discrepancy was defined as: 
\begin{equation}
   \epsilon_{\lambda_\textrm{i}} \equiv 100\cdot\dfrac{\left\|\lambda^*_\textrm{nm}- \lambda_{i}\right\|}{\lambda^*_\textrm{nm}}\ [\%]. 
\end{equation}

Finally, we used a Markov chain Monte Carlo approach to sample the parameters of each mode, estimating their distribution and uncertainties, selecting the most probable mode (i.e. the peak of the distribution).
All parameters and their respective values of this section are listed in the top part of Table~\ref{tab:inp_par}.

\subsection{Results}

The parsec-scale, helical jet structure of \C, consisting of two threads $T_1$ and $T_2$ is best approximated by the combination of four modes: a helical surface ($H_s$) and body ($H_{b1}$) mode and an elliptical surface ($E_s$) and body ($E_{b1}$) mode.
The resulting fit parameters are shown in Table \ref{tab:fit_amp}.
We find that the surface modes exhibit longer wavelengths than the body modes, in line with theoretical expectations \citep[e.g.][]{Hardee13}.
Their amplitudes also follow the same trend.

Determining the measurables $M_\textrm{x}$, $M_\textrm{j}$, as well as the jet pattern speed $\beta_\textrm{w}$, sound speed ($\alpha_\textrm{j}$), and ambient medium sound speed ($\alpha_\textrm{x}$) requires a relation between them and the measured K-H modes.
Such a relation exists between the resonant wavelength $\lambda^*$ \citep[see also][]{VegaGarcia19} and $M_\textrm{x}$, which is given by: 
\begin{equation}
 M_\textrm{x} = \frac{\lambda^*\beta_\textrm{j}\left(1 - \beta_\textrm{w}\cos \theta_\textrm{j}\right)}
 {8R_\textrm{j}\beta_\textrm{w}\sin\theta_\textrm{j}}.
 \label{eq:Mx}
\end{equation}
Here $\beta_\textrm{w}$ is given by:
\begin{equation}
   \beta_\textrm{w} = \frac{\beta_\textrm{w}^\textrm{app}}{\sin\theta_\textrm{j} + \beta_\textrm{w}^\textrm{app} \cos\theta_\textrm{j}}, \label{eq:bwapp}
\end{equation}
where we identified the apparent pattern velocity $\beta_\textrm{w}^\textrm{app}$ with that of component `C3' \citep{Nagai14}.
On the other hand, $M_\textrm{j}$ is given by:
\begin{equation}
    M_\textrm{j} = \frac{\lambda^*\left(1 - \beta_\textrm{w}\cos \theta_\textrm{j}\right)}{8R_\textrm{j}\Gamma_\textrm{j}\left(1-\beta_\textrm{w}/\beta_\textrm{j}\right)\sin\theta_\textrm{j}},
\end{equation}
and its relativistic counterpart $\mathcal{M}_\textrm{j}\equiv\Gamma_\textrm{j}M_\textrm{j}$.
Finally, the aforementioned sound speeds can then be calculated via the relation:
\begin{equation}
    \alpha_\textrm{[j,\ x]} = \beta_\textrm{j}/M_\textrm{[j,\ x]}. \label{eq:sound}
\end{equation}
In this framework, the derived jet values based on our instability analysis are listed in the bottom part of Table~\ref{tab:inp_par}.

\section{Discussion}\label{sec:Discussion}

\subsection{Modelled jet characteristics}

Our analysis presented in the previous section presents a self-consistent picture of the internal jet structure.
A strong indication in favour of our model is that we found the measured and predicted wavelengths to agree within the error budget.
The associated $M_\textrm{j, x}$ have moderate values with $M_\textrm{x} > M_\textrm{j}$, indicating that jet's density is lower than that of its surrounding interstellar medium.
This is consistent with the reported internal ($n_\textrm{e}^\textrm{int}$) and external ($n_\textrm{e}^\textrm{ext}$) density values for \C.
Specifically, \cite{Kim19} report an $n_\textrm{e}^\textrm{int}=0.6\pm0.4\,\textrm{cm}^{-3}$, whereas a number of estimates exist for $n_\textrm{e}^\textrm{ext}\sim10^1-10^5\,\textrm{cm}^{-3}$ \citep[e.g.][]{Fujita17, Nagai17, Park24}.
Furthermore, the sound speeds of the internal and external medium are consistent with a light, electron-positron plasma pair \citep{Lobanov01}.
Interestingly, a VLBI-independent approach by \cite{Paraschos23} using variability light curves came to the same conclusion about the \C\ jet's particle composition.
The low value of $\alpha_\textrm{x}$ points to a sub-relativistic ambient medium.
Finally, the two threads also naturally explain the prominent limb brightening of \C\ as a manifestation of the K-H instability.

\begin{table}[h!]
\centering
\caption{Predicted K-H mode resonant and longest unstable wavelengths}
\begin{tabular}{c|c|c}
\hline
\hline
K-H mode & $\lambda^*_{\textrm{pred};n,m}$ (mas) & $\lambda^1_{n,m}$ (mas) \\
\hline
$H_s$    & $6.7\pm0.8$  & $19.0\pm9.0$ \\ %
$E_s$    & $3.5\pm0.2$  & $6.2\pm3.0$ \\  %
$H_{b1}$ & $2.5\pm0.1$  & $3.7\pm1.8$ \\  %
$E_{b1}$ & $2.0\pm0.1$  & $2.7\pm1.3$ \\  %
\hline
\end{tabular}
\label{tab:pred}
\end{table}

\subsection{Brightness enhancements}\label{ssec:Spots}

Numerous jets, across different spatial scales are characterised by moving components, the trajectories of which have been traced for years \citep[e.g.][]{Weaver22}.
Theories have been put forth to explain them; for example, they have been attributed to Doppler boosting due to geometric effects, or shocks propagating through the bulk jet flow \citep[see e.g.][]{Liodakis20, Paraschos25}.
Their exact nature, though, still remains elusive.

The sub-parsec scale jet in \C\ is similarly characterised by such emission features in its downstream region.
In Fig.~\ref{fig:threads} we can identify one such brightness enhancement in the south-west of the core, at a distance of $\sim2$\,mas.
Interestingly, this area coincides with an overlap of the two threads.
As discussed in \cite{Nikonov23} for the M\,87 jet, such crossing K-H threads can also provide a natural explanation of regions of enhanced emission.
There, the authors also identify a spectral index flattening in the area where the crossing occurs.
A spectral index map (22-43\,GHz) of \C, characterised by similar resolution to this work, was presented in \cite{Park24} in observations taken a year later. 
Intriguingly, the spectral index appears flatter in that area as well\footnote{We note, however, that in \cite{Paraschos21} the spectral index appears more homogeneous in that area, which can be explained by source variability over longer timescales.}, while also being brighter than the rest of the approaching jet.
In combination with our work, this provides further evidence for overlapping K-H threads being a plausible mechanism of brightness enhancements in jets, in the optically thin regime.

\subsection{Driving mechanism}\label{ssec:Modelling}

Connecting the predicted longest unstable wavelengths $\lambda^1_{n, m}$ to an external driving mechanism requires knowledge about the mechanism's driving period.
Furthermore, the body modes will be attenuated at the longest unstable wavelength $\lambda^1_{n, m}$, such that:
\begin{equation}
\lambda^1_{n, m} \lesssim 2\lambda^*_{\textrm{pred};n,m}    \label{eq:l1}
\end{equation}
where $\lambda^*_{\textrm{pred};n,m}$ is the predicted resonant wavelength per K-H mode.
On the other hand, body modes are characterised by the absence of such a longest unstable wavelength and can theoretically have arbitrarily long wavelengths \citep{Lobanov01}.
The $\lambda^*_{\textrm{pred};n,m}$ values are shown in Table~\ref{tab:pred}.

A number of different periodicity values have been identified in \C, based on different measurables, which differ by orders of magnitude.
For example, \cite{Britzen19} identified a periodicity of the order of 40-100 years, based on radio light-curves.
Similarly, \cite{Dominik21} provide a periodicity estimate of up to 30 years, based on the movement of the jet internal position angle.
On the other hand, X-ray observations of the same source are suggestive of time scales of the order of tens of millions of years \cite{Dunn06}.
Using then the predicted $\lambda^*_{\textrm{pred};n,m}$ from our K-H analysis we can calculate a combination of the periodicity $T_\textrm{p}$ and the longest unstable wavelength $\lambda^1$, which satisfies the constraint imposed by Eq.~\ref{eq:l1}.
This is achieved by using the relation which links $T_\textrm{p}$ to $\lambda^1$ \citep[see][]{Lobanov01} and is given by:
\begin{equation}
    T_\textrm{p} = \frac{\left(1+z\right)\lambda^1\left(1+\eta_\textrm{j}^{0.5}/\delta_\textrm{j}\right)}{\beta_\textrm{j}\delta_\textrm{j}\sin\theta_\textrm{j}\left(1+\eta_\textrm{j}^{0.5}\Gamma_\textrm{j}\right)}
\end{equation}
Here $\delta_\textrm{j}$ is the jet Doppler factor, such that $\delta_\textrm{j}\equiv(\Gamma_\textrm{j}(1-\beta_\textrm{j}\sin\theta_\textrm{j}))^{-1}$.
We find that, in order to satisfy Eq.~\ref{eq:l1}, a periodicity of $T_\textrm{p}=\tp$ is required, resulting in the predictions for $\lambda^1_{n,m}$ shown in Table~\ref{tab:pred}.
Our computed periodicity is more consistent with the first two literature values \citep[see][]{Britzen19, Dominik21}, which are associated with jet precession.
Finally, we compared our computed periodicity to the structural changes of the \C\ jet as shown in \cite{Kam24}.
There the authors display the spatial evolution of the jet structure over a 12 year time span, which roughly equals half of $T_\textrm{p}$.
Based on our prediction, we would expect that the sinusoidal nature of the overall jet structure, imposed by the helical surface mode, has shifted the jet direction by half a wavelength.
Intriguingly, this behaviour is exactly mirrored in images of \cite{Kam24}.
The inner jet starts out in 2010 moving in south-eastward direction and by 2022 is moving south-westward.
Consequently, we predict that after another $\sim T_\textrm{p}/2$ the inner jet will again be directed in the south-eastward direction.

Furthermore, we contrast our result to the analysis of the blazar 3C\,345, presented in \cite{Lobanov05b}.
In their work, they calculate characteristic time scales for accretion disc activity in a binary system, such as thermal processes in the accretion disc ($\sim10$ years), the rotational ($\sim200$ years) and precession ($\sim2500$ years) period of the accretion disc, and the orbital ($\sim500$ years) period.
Their estimation of that source's black hole mass ($M^\textrm{3C\,345}_\bullet\sim1.4\times10^9\,\textrm{M}_\odot$) is similar to that of the \C\ system ($M^\textrm{\C}_\bullet\sim2\times10^9\,\textrm{M}_\odot$; \citealt{Giovannini18}), allowing for a direct comparison.
Our estimate for $T_\textrm{p}$ is in this case more aligned with effects due to disc activity.

\subsection{Current driven instability}

While K-H instabilities have been shown to be able to survive throughout the acceleration and collimation zone and in a highly magnetised regime \citep[see e.g.][]{Nikonov23}, it is worth to also consider the effect of a CDI (kink instability) on the \C\ jet.
For such an instability to form, the Kruskal-Shafranov criterion \citep{Bateman78} must be satisfied in the ideal magneto-hydrodynamic limit.
Specifically, the toroidal component of the magnetic field ($B_{\upphi}$) must dominate the poloidal ($B_\textrm{p}$) one, such that:
\begin{equation}
    q = \frac{2\uppi R_\textrm{j}B_\textrm{p}}{LB_{\upphi}} < 1, \label{eq:KS}
\end{equation}
where $L$ is the length of the jet.
For $L\sim6\,\textrm{mas}$ in our case and since $B_{\upphi}>B_\textrm{p}$ \citep{Paraschos24b}, the $q<1$ condition is satisfied.
Therefore, a CDI could potentially grow in the conditions identified in the \C\ jet.
At this point it is important to note that, as highlighted in \cite{Begelman98}, current-driven instabilities tend to lead to jet disruption and magnetic reconnection rather than simple wave-like patterns.
\C, however, is known for harbouring an ordered magnetic field \citep{Paraschos24} and is also exhibiting such wave-like patterns, as shown in this work.

Overall, there is no simple analytic approach to reliably evaluate the evolution or observational impact of a CDI scenario \citep{Begelman98,Bodo13}. 
To fully evaluate the growth and potential observational signatures of CDIs, numerical simulations are required -- an analysis that lies beyond the scope of this work and will be addressed in a future study.

\section{Conclusions} \label{sec:Conclusions}

In this work we presented an exploration of jet morphology of \C\ via K-H instability modelling.
Our analysis can be summarised as follows:
\begin{itemize}
    \item We modelled the sub-parsec jet structure of \C\ by fitting a bimodal function to slices drawn transversely to the bulk jet flow.
    \item By categorising each of the two modelled peaks based on its respective amplitude, we were able to retrieve an intertwining, helical morphology.
    \item We found that a Kelvin-Helmholtz instability can adequately describe this morphology; the corresponding Mach numbers and sound speeds for the internal (external) medium are $M_\textrm{j} = \Mj,\ \alpha_\textrm{j} = \ajj$ ($M_\textrm{x} = \Mx,\ \alpha_\textrm{x} = \ax$).
    \item These values are indicative of a low density, electron-positron pair plasma flow, surrounded by a sub-relativistic ambient medium.
    \item The timescales associated with these values favour the underlying periodic driving mechanism to be connected to accretion disc related processes. 
\end{itemize}

In summary, we have shown that high sensitivity, high fidelity images of nearby jetted-AGN can be utilised to explore the interplay between outflows and instabilities.
Future facilities, such as the next generation Very Large Array and the Square Kilometre Array, will offer even more advancements in sensitivity and will, thus, enable us to increase the number of jets, in which such an analysis is possible.

\begin{acknowledgements}
      We would like to thank the anonymous referee for the beneficial comments.
      {We would also like to thank I. Liodakis, A. Lobanov, A. Nikonov, and T. Savolainen for the fruitful discussions, which improved this manuscript greatly.}
      This research is supported by the European Research Council advanced grant “M2FINDERS - Mapping Magnetic Fields with INterferometry Down to Event hoRizon Scales” (Grant No. 101018682). 
      The European VLBI Network is a joint facility of independent European, African, Asian, and North American radio astronomy institutes.
      Scientific results from data presented in this publication are derived from the following EVN project code: GP058.
      This research has made use of the NASA/IPAC Extragalactic Database (NED), which is operated by the Jet Propulsion Laboratory, California Institute of Technology, under contract with the National Aeronautics and Space Administration. 
      This research has also made use of NASA's Astrophysics Data System Bibliographic Services. 
      Finally, this research made use of the following python packages: {\it numpy} \citep{Harris20}, {\it scipy} \citep{2020SciPy-NMeth}, {\it matplotlib} \citep{Hunter07}, {\it astropy} \citep{2013A&A...558A..33A, 2018AJ....156..123A} and {\it Uncertainties: a Python package for calculations with uncertainties.
      }
\end{acknowledgements}

\bibliographystyle{aa} 
\bibliography{aanda}

\begin{thebibliography}{62}
\expandafter\ifx\csname natexlab\endcsname\relax\def\natexlab#1{#1}\fi

\bibitem[{{Astropy Collaboration} {et~al.}(2018){Astropy Collaboration},
  {Price-Whelan}, {Sip{\H{o}}cz}, {G{\"u}nther}, {Lim}, {Crawford}, {Conseil},
  {Shupe}, {Craig}, {Dencheva}, {Ginsburg}, {VanderPlas}, {Bradley},
  {P{\'e}rez-Su{\'a}rez}, {de Val-Borro}, {Aldcroft}, {Cruz}, {Robitaille},
  {Tollerud}, {Ardelean}, {Babej}, {Bach}, {Bachetti}, {Bakanov}, {Bamford},
  {Barentsen}, {Barmby}, {Baumbach}, {Berry}, {Biscani}, {Boquien}, {Bostroem},
  {Bouma}, {Brammer}, {Bray}, {Breytenbach}, {Buddelmeijer}, {Burke},
  {Calderone}, {Cano Rodr{\'\i}guez}, {Cara}, {Cardoso}, {Cheedella}, {Copin},
  {Corrales}, {Crichton}, {D'Avella}, {Deil}, {Depagne}, {Dietrich}, {Donath},
  {Droettboom}, {Earl}, {Erben}, {Fabbro}, {Ferreira}, {Finethy}, {Fox},
  {Garrison}, {Gibbons}, {Goldstein}, {Gommers}, {Greco}, {Greenfield},
  {Groener}, {Grollier}, {Hagen}, {Hirst}, {Homeier}, {Horton}, {Hosseinzadeh},
  {Hu}, {Hunkeler}, {Ivezi{\'c}}, {Jain}, {Jenness}, {Kanarek}, {Kendrew},
  {Kern}, {Kerzendorf}, {Khvalko}, {King}, {Kirkby}, {Kulkarni}, {Kumar},
  {Lee}, {Lenz}, {Littlefair}, {Ma}, {Macleod}, {Mastropietro}, {McCully},
  {Montagnac}, {Morris}, {Mueller}, {Mumford}, {Muna}, {Murphy}, {Nelson},
  {Nguyen}, {Ninan}, {N{\"o}the}, {Ogaz}, {Oh}, {Parejko}, {Parley}, {Pascual},
  {Patil}, {Patil}, {Plunkett}, {Prochaska}, {Rastogi}, {Reddy Janga},
  {Sabater}, {Sakurikar}, {Seifert}, {Sherbert}, {Sherwood-Taylor}, {Shih},
  {Sick}, {Silbiger}, {Singanamalla}, {Singer}, {Sladen}, {Sooley},
  {Sornarajah}, {Streicher}, {Teuben}, {Thomas}, {Tremblay}, {Turner},
  {Terr{\'o}n}, {van Kerkwijk}, {de la Vega}, {Watkins}, {Weaver}, {Whitmore},
  {Woillez}, {Zabalza}, \& {Astropy Contributors}}]{2018AJ....156..123A}
{Astropy Collaboration}, {Price-Whelan}, A.~M., {Sip{\H{o}}cz}, B.~M., {et~al.}
  2018, \aj, 156, 123

\bibitem[{{Astropy Collaboration} {et~al.}(2013){Astropy Collaboration},
  {Robitaille}, {Tollerud}, {Greenfield}, {Droettboom}, {Bray}, {Aldcroft},
  {Davis}, {Ginsburg}, {Price-Whelan}, {Kerzendorf}, {Conley}, {Crighton},
  {Barbary}, {Muna}, {Ferguson}, {Grollier}, {Parikh}, {Nair}, {Unther},
  {Deil}, {Woillez}, {Conseil}, {Kramer}, {Turner}, {Singer}, {Fox}, {Weaver},
  {Zabalza}, {Edwards}, {Azalee Bostroem}, {Burke}, {Casey}, {Crawford},
  {Dencheva}, {Ely}, {Jenness}, {Labrie}, {Lim}, {Pierfederici}, {Pontzen},
  {Ptak}, {Refsdal}, {Servillat}, \& {Streicher}}]{2013A&A...558A..33A}
{Astropy Collaboration}, {Robitaille}, T.~P., {Tollerud}, E.~J., {et~al.} 2013,
  \aap, 558, A33

\bibitem[{{Bateman}(1978)}]{Bateman78}
{Bateman}, G. 1978, {MHD instabilities}

\bibitem[{{Begelman}(1998)}]{Begelman98}
{Begelman}, M.~C. 1998, \apj, 493, 291

\bibitem[{{Bodo} {et~al.}(2013){Bodo}, {Mamatsashvili}, {Rossi}, \&
  {Mignone}}]{Bodo13}
{Bodo}, G., {Mamatsashvili}, G., {Rossi}, P., \& {Mignone}, A. 2013, \mnras,
  434, 3030

\bibitem[{{Britzen} {et~al.}(2019){Britzen}, {Fendt}, {Zaja{\v{c}}ek}, {Jaron},
  {Pashchenko}, {Aller}, \& {Aller}}]{Britzen19}
{Britzen}, S., {Fendt}, C., {Zaja{\v{c}}ek}, M., {et~al.} 2019, Galaxies, 7, 72

\bibitem[{{Britzen} {et~al.}(2024){Britzen}, {Kova{\v{c}}evi{\'c}},
  {Zaja{\v{c}}ek}, {Popovi{\'c}}, {Pashchenko}, {Kun}, {P{\'a}nis}, {Jaron},
  {Pl{\v{s}}ek}, {Tursunov}, \& {Stuchl{\'\i}k}}]{Britzen24}
{Britzen}, S., {Kova{\v{c}}evi{\'c}}, A.~B., {Zaja{\v{c}}ek}, M., {et~al.}
  2024, \mnras [\eprint[arXiv]{2410.18184}]

\bibitem[{{Dominik} {et~al.}(2021){Dominik}, {Linhoff}, {Els{\"a}sser}, \&
  {Rhode}}]{Dominik21}
{Dominik}, R.~M., {Linhoff}, L., {Els{\"a}sser}, D., \& {Rhode}, W. 2021,
  \mnras, 503, 5448

\bibitem[{{Dunn} {et~al.}(2006){Dunn}, {Fabian}, \& {Sanders}}]{Dunn06}
{Dunn}, R.~J.~H., {Fabian}, A.~C., \& {Sanders}, J.~S. 2006, \mnras, 366, 758

\bibitem[{{Fuentes} {et~al.}(2018){Fuentes}, {G{\'o}mez}, {Mart{\'\i}}, \&
  {Perucho}}]{Fuentes18}
{Fuentes}, A., {G{\'o}mez}, J.~L., {Mart{\'\i}}, J.~M., \& {Perucho}, M. 2018,
  \apj, 860, 121

\bibitem[{{Fuentes} {et~al.}(2023){Fuentes}, {G{\'o}mez}, {Mart{\'\i}},
  {Perucho}, {Zhao}, {Lico}, {Lobanov}, {Bruni}, {Kovalev}, {Chael}, {Akiyama},
  {Bouman}, {Sun}, {Cho}, {Traianou}, {Toscano}, {Dahale}, {Foschi}, {Gurvits},
  {Jorstad}, {Kim}, {Marscher}, {Mizuno}, {Ros}, \& {Savolainen}}]{Fuentes23}
{Fuentes}, A., {G{\'o}mez}, J.~L., {Mart{\'\i}}, J.~M., {et~al.} 2023, Nature
  Astronomy, 7, 1359

\bibitem[{{Fujita} \& {Nagai}(2017)}]{Fujita17}
{Fujita}, Y. \& {Nagai}, H. 2017, \mnras, 465, L94

\bibitem[{{Giovannini} {et~al.}(2018){Giovannini}, {Savolainen}, {Orienti},
  {Nakamura}, {Nagai}, {Kino}, {Giroletti}, {Hada}, {Bruni}, {Kovalev},
  {Anderson}, {D'Ammando}, {Hodgson}, {Honma}, {Krichbaum}, {Lee}, {Lico},
  {Lisakov}, {Lobanov}, {Petrov}, {Sohn}, {Sokolovsky}, {Voitsik}, {Zensus}, \&
  {Tingay}}]{Giovannini18}
{Giovannini}, G., {Savolainen}, T., {Orienti}, M., {et~al.} 2018, Nature
  Astronomy, 2, 472

\bibitem[{{Hardee}(1984)}]{Hardee84}
{Hardee}, P.~E. 1984, \apj, 287, 523

\bibitem[{{Hardee}(1986)}]{Hardee86}
{Hardee}, P.~E. 1986, \apj, 303, 111

\bibitem[{{Hardee}(1987)}]{Hardee87}
{Hardee}, P.~E. 1987, \apj, 313, 607

\bibitem[{{Hardee}(2000)}]{Hardee00}
{Hardee}, P.~E. 2000, \apj, 533, 176

\bibitem[{{Hardee}(2003)}]{Hardee03}
{Hardee}, P.~E. 2003, \apj, 597, 798

\bibitem[{{Hardee}(2013)}]{Hardee13}
{Hardee}, P.~E. 2013, in European Physical Journal Web of Conferences, Vol.~61,
  European Physical Journal Web of Conferences, 02001

\bibitem[{{Hardee} {et~al.}(1997){Hardee}, {Clarke}, \& {Rosen}}]{Hardee97}
{Hardee}, P.~E., {Clarke}, D.~A., \& {Rosen}, A. 1997, \apj, 485, 533

\bibitem[{{Hardee} \& {Eilek}(2011)}]{Hardee11}
{Hardee}, P.~E. \& {Eilek}, J.~A. 2011, \apj, 735, 61

\bibitem[{Harris {et~al.}(2020)Harris, Millman, van~der Walt, Gommers,
  Virtanen, Cournapeau, Wieser, Taylor, Berg, Smith, Kern, Picus, Hoyer, van
  Kerkwijk, Brett, Haldane, del R{'{\i}}o, Wiebe, Peterson,
  G{'{e}}rard-Marchant, Sheppard, Reddy, Weckesser, Abbasi, Gohlke, \&
  Oliphant}]{Harris20}
Harris, C.~R., Millman, K.~J., van~der Walt, S.~J., {et~al.} 2020, Nature, 585,
  357

\bibitem[{{Hodgson} {et~al.}(2018){Hodgson}, {Rani}, {Lee}, {Algaba}, {Kino},
  {Trippe}, {Park}, {Zhao}, {Byun}, {Kang}, {Kim}, {Kim}, {Kim}, {Miyazaki},
  {Wajima}, {Oh}, {Kim}, \& {Gurwell}}]{Hodgson18}
{Hodgson}, J.~A., {Rani}, B., {Lee}, S.-S., {et~al.} 2018, \mnras, 475, 368

\bibitem[{{Hodgson} {et~al.}(2021){Hodgson}, {Rani}, {Oh}, {Marscher},
  {Jorstad}, {Mizuno}, {Park}, {Lee}, {Trippe}, \& {Mertens}}]{Hodgson21}
{Hodgson}, J.~A., {Rani}, B., {Oh}, J., {et~al.} 2021, \apj, 914, 43

\bibitem[{Hunter(2007)}]{Hunter07}
Hunter, J.~D. 2007, Computing in Science \& Engineering, 9, 90

\bibitem[{{Janssen} {et~al.}(2019){Janssen}, {Goddi}, {van Bemmel}, {Kettenis},
  {Small}, {Liuzzo}, {Rygl}, {Mart{\'\i}-Vidal}, {Blackburn}, {Wielgus}, \&
  {Falcke}}]{Janssen19}
{Janssen}, M., {Goddi}, C., {van Bemmel}, I.~M., {et~al.} 2019, \aap, 626, A75

\bibitem[{{Kam} {et~al.}(2024){Kam}, {Hodgson}, {Park}, {Kino}, {Nagai},
  {Trippe}, \& {Wagner}}]{Kam24}
{Kam}, M., {Hodgson}, J.~A., {Park}, J., {et~al.} 2024, \apj, 970, 176

\bibitem[{{Kim} {et~al.}(2019){Kim}, {Krichbaum}, {Marscher}, {Jorstad},
  {Agudo}, {Thum}, {Hodgson}, {MacDonald}, {Ros}, {Lu}, {Bremer}, {de Vicente},
  {Lindqvist}, {Trippe}, \& {Zensus}}]{Kim19}
{Kim}, J.~Y., {Krichbaum}, T.~P., {Marscher}, A.~P., {et~al.} 2019, \aap, 622,
  A196

\bibitem[{{Kramer} {et~al.}(2024){Kramer}, {MacDonald}, {Paraschos}, \&
  {Ricci}}]{Kramer24}
{Kramer}, J.~A., {MacDonald}, N.~R., {Paraschos}, G.~F., \& {Ricci}, L. 2024,
  \aap, 691, A14

\bibitem[{{Liodakis} {et~al.}(2020){Liodakis}, {Blinov}, {Jorstad}, {Arkharov},
  {Di Paola}, {Efimova}, {Grishina}, {Kiehlmann}, {Kopatskaya}, {Larionov},
  {Larionova}, {Larionova}, {Marscher}, {Morozova}, {Nikiforova}, {Pavlidou},
  {Traianou}, {Troitskaya}, {Troitsky}, {Uemura}, \& {Weaver}}]{Liodakis20}
{Liodakis}, I., {Blinov}, D., {Jorstad}, S.~G., {et~al.} 2020, \apj, 902, 61

\bibitem[{{Lobanov} {et~al.}(2003){Lobanov}, {Hardee}, \& {Eilek}}]{Lobanov03}
{Lobanov}, A., {Hardee}, P., \& {Eilek}, J. 2003, \nar, 47, 629

\bibitem[{{Lobanov} \& {Roland}(2005)}]{Lobanov05b}
{Lobanov}, A.~P. \& {Roland}, J. 2005, \aap, 431, 831

\bibitem[{{Lobanov} \& {Zensus}(2001)}]{Lobanov01}
{Lobanov}, A.~P. \& {Zensus}, J.~A. 2001, Science, 294, 128

\bibitem[{{Mart{\'\i}-Vidal} {et~al.}(2021){Mart{\'\i}-Vidal}, {Mus},
  {Janssen}, {de Vicente}, \& {Gonz{\'a}lez}}]{MartiVidal21}
{Mart{\'\i}-Vidal}, I., {Mus}, A., {Janssen}, M., {de Vicente}, P., \&
  {Gonz{\'a}lez}, J. 2021, \aap, 646, A52

\bibitem[{{McKinney} \& {Blandford}(2009)}]{McKinney09}
{McKinney}, J.~C. \& {Blandford}, R.~D. 2009, \mnras, 394, L126

\bibitem[{{Mizuno} {et~al.}(2015){Mizuno}, {G{\'o}mez}, {Nishikawa}, {Meli},
  {Hardee}, \& {Rezzolla}}]{Mizuno15}
{Mizuno}, Y., {G{\'o}mez}, J.~L., {Nishikawa}, K.-I., {et~al.} 2015, \apj, 809,
  38

\bibitem[{{Mizuno} {et~al.}(2007){Mizuno}, {Hardee}, \& {Nishikawa}}]{Mizuno07}
{Mizuno}, Y., {Hardee}, P., \& {Nishikawa}, K.-I. 2007, \apj, 662, 835

\bibitem[{{Mizuno} {et~al.}(2009){Mizuno}, {Lyubarsky}, {Nishikawa}, \&
  {Hardee}}]{Mizuno09}
{Mizuno}, Y., {Lyubarsky}, Y., {Nishikawa}, K.-I., \& {Hardee}, P.~E. 2009,
  \apj, 700, 684

\bibitem[{{Nagai} {et~al.}(2016){Nagai}, {Chida}, {Kino}, {Orienti},
  {D'Ammando}, {Giovannini}, \& {Hiura}}]{Nagai16}
{Nagai}, H., {Chida}, H., {Kino}, M., {et~al.} 2016, Astronomische Nachrichten,
  337, 69

\bibitem[{{Nagai} {et~al.}(2017){Nagai}, {Fujita}, {Nakamura}, {Orienti},
  {Kino}, {Asada}, \& {Giovannini}}]{Nagai17}
{Nagai}, H., {Fujita}, Y., {Nakamura}, M., {et~al.} 2017, \apj, 849, 52

\bibitem[{{Nagai} {et~al.}(2014){Nagai}, {Haga}, {Giovannini}, {Doi},
  {Orienti}, {D'Ammando}, {Kino}, {Nakamura}, {Asada}, {Hada}, \&
  {Giroletti}}]{Nagai14}
{Nagai}, H., {Haga}, T., {Giovannini}, G., {et~al.} 2014, \apj, 785, 53

\bibitem[{{Nikonov} {et~al.}(2023){Nikonov}, {Kovalev}, {Kravchenko},
  {Pashchenko}, \& {Lobanov}}]{Nikonov23}
{Nikonov}, A.~S., {Kovalev}, Y.~Y., {Kravchenko}, E.~V., {Pashchenko}, I.~N.,
  \& {Lobanov}, A.~P. 2023, \mnras, 526, 5949

\bibitem[{{Oh} {et~al.}(2022){Oh}, {Hodgson}, {Trippe}, {Krichbaum}, {Kam},
  {Paraschos}, {Kim}, {Rani}, {Sohn}, {Lee}, {Lico}, {Liuzzo}, {Bremer}, \&
  {Zensus}}]{Oh22}
{Oh}, J., {Hodgson}, J.~A., {Trippe}, S., {et~al.} 2022, \mnras, 509, 1024

\bibitem[{{Paraschos}(2025)}]{Paraschos25}
{Paraschos}, G.~F. 2025, \aap, 695, L3

\bibitem[{{Paraschos} {et~al.}(2024{\natexlab{a}}){Paraschos}, {Debbrecht},
  {Kramer}, {Traianou}, {Liodakis}, {Krichbaum}, {Kim}, {Janssen}, {Nair},
  {Savolainen}, {Ros}, {Bach}, {Hodgson}, {Lisakov}, {MacDonald}, \&
  {Zensus}}]{Paraschos24b}
{Paraschos}, G.~F., {Debbrecht}, L.~C., {Kramer}, J.~A., {et~al.}
  2024{\natexlab{a}}, \aap, 686, L5

\bibitem[{{Paraschos} {et~al.}(2021){Paraschos}, {Kim}, {Krichbaum}, \&
  {Zensus}}]{Paraschos21}
{Paraschos}, G.~F., {Kim}, J.~Y., {Krichbaum}, T.~P., \& {Zensus}, J.~A. 2021,
  \aap, 650, L18

\bibitem[{{Paraschos} {et~al.}(2024{\natexlab{b}}){Paraschos}, {Kim},
  {Wielgus}, {R{\"o}der}, {Krichbaum}, {Ros}, {Agudo}, {Myserlis},
  {Moscibrodzka}, {Traianou}, {Zensus}, {Blackburn}, {Chan}, {Issaoun},
  {Janssen}, {Johnson}, {Fish}, {Akiyama}, {Alberdi}, {Alef}, {Algaba},
  {Anantua}, {Asada}, {Azulay}, {Bach}, {Baczko}, {Ball}, {Balokovi{\'c}},
  {Barrett}, {Baub{\"o}ck}, {Benson}, {Bintley}, {Blundell}, {Bouman}, {Bower},
  {Boyce}, {Bremer}, {Brinkerink}, {Brissenden}, {Britzen}, {Broderick},
  {Broguiere}, {Bronzwaer}, {Bustamante}, {Byun}, {Carlstrom}, {Ceccobello},
  {Chael}, {Chang}, {Chatterjee}, {Chatterjee}, {Chen}, {Chen}, {Cheng}, {Cho},
  {Christian}, {Conroy}, {Conway}, {Cordes}, {Crawford}, {Crew}, {Cruz-Osorio},
  {Cui}, {Dahale}, {Davelaar}, {De Laurentis}, {Deane}, {Dempsey}, {Desvignes},
  {Dexter}, {Dhruv}, {Doeleman}, {Dougal}, {Dzib}, {Eatough}, {Emami},
  {Falcke}, {Farah}, {Fomalont}, {Ford}, {Foschi}, {Fraga-Encinas}, {Freeman},
  {Friberg}, {Fromm}, {Fuentes}, {Galison}, {Gammie}, {Garc{\'\i}a}, {Gentaz},
  {Georgiev}, {Goddi}, {Gold}, {G{\'o}mez-Ruiz}, {G{\'o}mez}, {Gu}, {Gurwell},
  {Hada}, {Haggard}, {Haworth}, {Hecht}, {Hesper}, {Heumann}, {Ho}, {Ho},
  {Honma}, {Huang}, {Huang}, {Hughes}, {Ikeda}, {Impellizzeri}, {Inoue},
  {James}, {Jannuzi}, {Jeter}, {Jaing}, {Jim{\'e}nez-Rosales}, {Jorstad},
  {Joshi}, {Jung}, {Karami}, {Karuppusamy}, {Kawashima}, {Keating}, {Kettenis},
  {Kim}, {Kim}, {Kim}, {Kino}, {Koay}, {Kocherlakota}, {Kofuji}, {Koch},
  {Koyama}, {Kramer}, {Kramer}, {Kramer}, {Kuo}, {La Bella}, {Lauer}, {Lee},
  {Lee}, {Leung}, {Levis}, {Li}, {Lico}, {Lindahl}, {Lindqvist}, {Lisakov},
  {Liu}, {Liu}, {Liuzzo}, {Lo}, {Lobanov}, {Loinard}, {Lonsdale}, {Lowitz},
  {Lu}, {MacDonald}, {Mao}, {Marchili}, {Markoff}, {Marrone}, {Marscher},
  {Mart{\'\i}-Vidal}, {Matsushita}, {Matthews}, {Medeiros}, {Menten},
  {Michalik}, {Mizuno}, {Mizuno}, {Moran}, {Moriyama}, {Mulaudzi},
  {M{\"u}ller}, {M{\"u}ller}, {Mus}, {Musoke}, {Nadolski}, {Nagai}, {Nagar},
  {Nakamura}, {Narayanan}, {Natarajan}, {Nathanail}, {Navarro Fuentes},
  {Neilsen}, {Neri}, {Ni}, {Noutsos}, {Nowak}, {Oh}, {Okino}, {Olivares},
  {Ortiz-Le{\'o}n}, {Oyama}, {{\"O}zel}, {Palumbo}, {Park}, {Parsons}, {Patel},
  {Pen}, {Pi{\'e}tu}, {Plambeck}, {PopStefanija}, {Porth}, {P{\"o}tzl},
  {Prather}, {Preciado-L{\'o}pez}, {Psaltis}, {Pu}, {Ramakrishnan}, {Rao},
  {Rawlings}, {Raymond}, {Rezzolla}, {Ricarte}, {Ripperda}, {Roelofs},
  {Rogers}, {Romero-Ca{\~n}izales}, {Roshanineshat}, {Rottmann}, {Roy}, {Ruiz},
  {Ruszczyk}, {Rygl}, {S{\'a}nchez}, {S{\'a}nchez-Arg{\"u}elles},
  {S{\'a}nchez-Portal}, {Sasada}, {Satapathy}, {Savolainen}, {Schloerb},
  {Schonfeld}, {Schuster}, {Shao}, {Shen}, {Small}, {Sohn}, {SooHoo},
  {Sosapanta Salas}, {Souccar}, {Sun}, {Tazaki}, {Tetarenko}, {Tiede},
  {Tilanus}, {Titus}, {Torne}, {Toscano}, {Trent}, {Trippe}, {Turk}, {van
  Bemmel}, {van Langevelde}, {van Rossum}, {Vos}, {Wagner}, {Ward-Thompson},
  {Wardle}, {Washington}, {Weintroub}, {Wharton}, {Wiik}, {Witzel}, {Wondrak},
  {Wong}, {Wu}, {Yadlapalli}, {Yamaguchi}, {Yfantis}, {Yoon}, {Young}, {Young},
  {Younsi}, {Yu}, {Yuan}, {Yuan}, {Zhang}, {Zhao}, \& {Zhao}}]{Paraschos24}
{Paraschos}, G.~F., {Kim}, J.~Y., {Wielgus}, M., {et~al.} 2024{\natexlab{b}},
  \aap, 682, L3

\bibitem[{{Paraschos} {et~al.}(2022){Paraschos}, {Krichbaum}, {Kim}, {Hodgson},
  {Oh}, {Ros}, {Zensus}, {Marscher}, {Jorstad}, {Gurwell},
  {L{\"a}hteenm{\"a}ki}, {Tornikoski}, {Kiehlmann}, \&
  {Readhead}}]{Paraschos22}
{Paraschos}, G.~F., {Krichbaum}, T.~P., {Kim}, J.~Y., {et~al.} 2022, \aap, 665,
  A1

\bibitem[{{Paraschos} {et~al.}(2023){Paraschos}, {Mpisketzis}, {Kim}, {Witzel},
  {Krichbaum}, {Zensus}, {Gurwell}, {L{\"a}hteenm{\"a}ki}, {Tornikoski},
  {Kiehlmann}, \& {Readhead}}]{Paraschos23}
{Paraschos}, G.~F., {Mpisketzis}, V., {Kim}, J.~Y., {et~al.} 2023, \aap, 669,
  A32

\bibitem[{{Park} {et~al.}(2024){Park}, {Kino}, {Nagai}, {Nakamura}, {Asada},
  {Kam}, \& {Hodgson}}]{Park24}
{Park}, J., {Kino}, M., {Nagai}, H., {et~al.} 2024, \aap, 685, A115

\bibitem[{{Pasetto} {et~al.}(2021){Pasetto}, {Carrasco-Gonz{\'a}lez},
  {G{\'o}mez}, {Mart{\'\i}}, {Perucho}, {O'Sullivan}, {Anderson},
  {D{\'\i}az-Gonz{\'a}lez}, {Fuentes}, \& {Wardle}}]{Pasetto21}
{Pasetto}, A., {Carrasco-Gonz{\'a}lez}, C., {G{\'o}mez}, J.~L., {et~al.} 2021,
  \apjl, 923, L5

\bibitem[{{Perucho} {et~al.}(2012){Perucho}, {Kovalev}, {Lobanov}, {Hardee}, \&
  {Agudo}}]{Perucho12}
{Perucho}, M., {Kovalev}, Y.~Y., {Lobanov}, A.~P., {Hardee}, P.~E., \& {Agudo},
  I. 2012, \apj, 749, 55

\bibitem[{{Perucho} {et~al.}(2005){Perucho}, {Mart{\'\i}}, \&
  {Hanasz}}]{Perucho05}
{Perucho}, M., {Mart{\'\i}}, J.~M., \& {Hanasz}, M. 2005, \aap, 443, 863

\bibitem[{{Planck Collaboration} {et~al.}(2016){Planck Collaboration}, {Ade},
  {Aghanim}, {Arnaud}, {Ashdown}, {Aumont}, {Baccigalupi}, {Banday},
  {Barreiro}, {Bartlett}, {Bartolo}, {Battaner}, {Battye}, {Benabed},
  {Beno{\^\i}t}, {Benoit-L{\'e}vy}, {Bernard}, {Bersanelli}, {Bielewicz},
  {Bock}, {Bonaldi}, {Bonavera}, {Bond}, {Borrill}, {Bouchet}, {Boulanger},
  {Bucher}, {Burigana}, {Butler}, {Calabrese}, {Cardoso}, {Catalano},
  {Challinor}, {Chamballu}, {Chary}, {Chiang}, {Chluba}, {Christensen},
  {Church}, {Clements}, {Colombi}, {Colombo}, {Combet}, {Coulais}, {Crill},
  {Curto}, {Cuttaia}, {Danese}, {Davies}, {Davis}, {de Bernardis}, {de Rosa},
  {de Zotti}, {Delabrouille}, {D{\'e}sert}, {Di Valentino}, {Dickinson},
  {Diego}, {Dolag}, {Dole}, {Donzelli}, {Dor{\'e}}, {Douspis}, {Ducout},
  {Dunkley}, {Dupac}, {Efstathiou}, {Elsner}, {En{\ss}lin}, {Eriksen},
  {Farhang}, {Fergusson}, {Finelli}, {Forni}, {Frailis}, {Fraisse},
  {Franceschi}, {Frejsel}, {Galeotta}, {Galli}, {Ganga}, {Gauthier}, {Gerbino},
  {Ghosh}, {Giard}, {Giraud-H{\'e}raud}, {Giusarma}, {Gjerl{\o}w},
  {Gonz{\'a}lez-Nuevo}, {G{\'o}rski}, {Gratton}, {Gregorio}, {Gruppuso},
  {Gudmundsson}, {Hamann}, {Hansen}, {Hanson}, {Harrison}, {Helou},
  {Henrot-Versill{\'e}}, {Hern{\'a}ndez-Monteagudo}, {Herranz}, {Hildebrandt},
  {Hivon}, {Hobson}, {Holmes}, {Hornstrup}, {Hovest}, {Huang}, {Huffenberger},
  {Hurier}, {Jaffe}, {Jaffe}, {Jones}, {Juvela}, {Keih{\"a}nen}, {Keskitalo},
  {Kisner}, {Kneissl}, {Knoche}, {Knox}, {Kunz}, {Kurki-Suonio}, {Lagache},
  {L{\"a}hteenm{\"a}ki}, {Lamarre}, {Lasenby}, {Lattanzi}, {Lawrence}, {Leahy},
  {Leonardi}, {Lesgourgues}, {Levrier}, {Lewis}, {Liguori}, {Lilje},
  {Linden-V{\o}rnle}, {L{\'o}pez-Caniego}, {Lubin}, {Mac{\'\i}as-P{\'e}rez},
  {Maggio}, {Maino}, {Mandolesi}, {Mangilli}, {Marchini}, {Maris}, {Martin},
  {Martinelli}, {Mart{\'\i}nez-Gonz{\'a}lez}, {Masi}, {Matarrese}, {McGehee},
  {Meinhold}, {Melchiorri}, {Melin}, {Mendes}, {Mennella}, {Migliaccio},
  {Millea}, {Mitra}, {Miville-Desch{\^e}nes}, {Moneti}, {Montier}, {Morgante},
  {Mortlock}, {Moss}, {Munshi}, {Murphy}, {Naselsky}, {Nati}, {Natoli},
  {Netterfield}, {N{\o}rgaard-Nielsen}, {Noviello}, {Novikov}, {Novikov},
  {Oxborrow}, {Paci}, {Pagano}, {Pajot}, {Paladini}, {Paoletti}, {Partridge},
  {Pasian}, {Patanchon}, {Pearson}, {Perdereau}, {Perotto}, {Perrotta},
  {Pettorino}, {Piacentini}, {Piat}, {Pierpaoli}, {Pietrobon}, {Plaszczynski},
  {Pointecouteau}, {Polenta}, {Popa}, {Pratt}, {Pr{\'e}zeau}, {Prunet},
  {Puget}, {Rachen}, {Reach}, {Rebolo}, {Reinecke}, {Remazeilles}, {Renault},
  {Renzi}, {Ristorcelli}, {Rocha}, {Rosset}, {Rossetti}, {Roudier},
  {Rouill{\'e} d'Orfeuil}, {Rowan-Robinson}, {Rubi{\~n}o-Mart{\'\i}n},
  {Rusholme}, {Said}, {Salvatelli}, {Salvati}, {Sandri}, {Santos},
  {Savelainen}, {Savini}, {Scott}, {Seiffert}, {Serra}, {Shellard}, {Spencer},
  {Spinelli}, {Stolyarov}, {Stompor}, {Sudiwala}, {Sunyaev}, {Sutton},
  {Suur-Uski}, {Sygnet}, {Tauber}, {Terenzi}, {Toffolatti}, {Tomasi},
  {Tristram}, {Trombetti}, {Tucci}, {Tuovinen}, {T{\"u}rler}, {Umana},
  {Valenziano}, {Valiviita}, {Van Tent}, {Vielva}, {Villa}, {Wade}, {Wandelt},
  {Wehus}, {White}, {White}, {Wilkinson}, {Yvon}, {Zacchei}, \&
  {Zonca}}]{Planck16}
{Planck Collaboration}, {Ade}, P.~A.~R., {Aghanim}, N., {et~al.} 2016, \aap,
  594, A13

\bibitem[{{Punsly} {et~al.}(2021){Punsly}, {Nagai}, {Savolainen}, \&
  {Orienti}}]{Punsly21}
{Punsly}, B., {Nagai}, H., {Savolainen}, T., \& {Orienti}, M. 2021, \apj, 911,
  19

\bibitem[{{Savolainen} {et~al.}(2023){Savolainen}, {Giovannini}, {Kovalev},
  {Perucho}, {Anderson}, {Bruni}, {Edwards}, {Fuentes}, {Giroletti},
  {G{\'o}mez}, {Hada}, {Lee}, {Lisakov}, {Lobanov}, {L{\'o}pez-Miralles},
  {Orienti}, {Petrov}, {Plavin}, {Sohn}, {Sokolovsky}, {Voitsik}, \&
  {Zensus}}]{Savolainen23}
{Savolainen}, T., {Giovannini}, G., {Kovalev}, Y.~Y., {et~al.} 2023, \aap, 676,
  A114

\bibitem[{{Shepherd} {et~al.}(1994){Shepherd}, {Pearson}, \&
  {Taylor}}]{Shepherd94}
{Shepherd}, M.~C., {Pearson}, T.~J., \& {Taylor}, G.~B. 1994, in Bulletin of
  the American Astronomical Society, Vol.~26, 987--989

\bibitem[{{Strauss} {et~al.}(1992){Strauss}, {Huchra}, {Davis}, {Yahil},
  {Fisher}, \& {Tonry}}]{Strauss92}
{Strauss}, M.~A., {Huchra}, J.~P., {Davis}, M., {et~al.} 1992, \apjs, 83, 29

\bibitem[{{Traianou} {et~al.}(2024){Traianou}, {Krichbaum}, {G{\'o}mez},
  {Lico}, {Paraschos}, {Cho}, {Ros}, {Zhao}, {Liodakis}, {Dahale}, {Toscano},
  {Fuentes}, {Foschi}, {Casadio}, {MacDonald}, {Kim}, {Hervet}, {Jorstad},
  {Lobanov}, {Hodgson}, {Myserlis}, {Agudo}, {Zensus}, \&
  {Marscher}}]{Traianou24}
{Traianou}, E., {Krichbaum}, T.~P., {G{\'o}mez}, J.~L., {et~al.} 2024, \aap,
  682, A154

\bibitem[{{Vega-Garc{\'\i}a} {et~al.}(2019){Vega-Garc{\'\i}a}, {Perucho}, \&
  {Lobanov}}]{VegaGarcia19}
{Vega-Garc{\'\i}a}, L., {Perucho}, M., \& {Lobanov}, A.~P. 2019, \aap, 627, A79

\bibitem[{Virtanen {et~al.}(2020)Virtanen, Gommers, Oliphant, Haberland, Reddy,
  Cournapeau, Burovski, Peterson, Weckesser, Bright, {van der Walt}, Brett,
  Wilson, Millman, Mayorov, Nelson, Jones, Kern, Larson, Carey, Polat, Feng,
  Moore, {VanderPlas}, Laxalde, Perktold, Cimrman, Henriksen, Quintero, Harris,
  Archibald, Ribeiro, Pedregosa, {van Mulbregt}, \& {SciPy 1.0
  Contributors}}]{2020SciPy-NMeth}
Virtanen, P., Gommers, R., Oliphant, T.~E., {et~al.} 2020, Nature Methods, 17,
  261

\bibitem[{{Weaver} {et~al.}(2022){Weaver}, {Jorstad}, {Marscher}, {Morozova},
  {Troitsky}, {Agudo}, {G{\'o}mez}, {L{\"a}hteenm{\"a}ki}, {Tammi}, \&
  {Tornikoski}}]{Weaver22}
{Weaver}, Z.~R., {Jorstad}, S.~G., {Marscher}, A.~P., {et~al.} 2022, \apjs,
  260, 12

\end{thebibliography}

\begin{appendix}
\section{Alternative thread identification}\label{app:Alt}

\begin{table*}[ht!]
\centering
\caption{Modelled parameters for alternative $\textrm{T}_1$ and $\textrm{T}_2$ identification}
\begin{tabular}{c|c|c|c|c|c|c|c}
\hline
\hline
Mode & $\lambda_\textrm{i}$ & $A_\textrm{i}$ & $\phi_\textrm{i}$ & $k_\textrm{i}$ & K-H & $\epsilon_{\lambda_\textrm{i}}$ & $\lambda^*_\textrm{n,m}$ \\ 
     &  \begin{tabular}[c]{@{}c@{}} (mas) \\ \hline \end{tabular}                &   \begin{tabular}[c]{@{}c@{}} (mas) \\ \hline \end{tabular}              &   \begin{tabular}[c]{@{}c@{}} (rad) \\ \hline \end{tabular}            &   \begin{tabular}[c]{@{}c@{}} (mas) \\ \hline \end{tabular}  &   \begin{tabular}[c]{@{}c@{}} mode \\ \hline \end{tabular}  &   \begin{tabular}[c]{@{}c@{}} {} \\ \hline \end{tabular}  &   \begin{tabular}[c]{@{}c@{}} (mas) \\ \hline \end{tabular} \\
     & T1 | T2              &  T1 | T2            &  T1 | T2          &  T1 | T2 &  T1 | T2 &  T1 | T2 &  T1 | T2 \\
\hline
1  & 6.3 | 7.1  & 0.4 | 2.0       & 4.0 | 6.3 & 0.4 & $H_\mathrm{s}$  & 4\%   | 18\%    & 10.0 \\ 
2  & 3.0 | 4.2  & 1.1 | 1.7       & 4.4 | 4.3 & 0.4 & $E_\mathrm{s}$  & 17\%  | 16\%    & 9.0  \\ 
3  & 2.3        & 0.5 | 1.5       & 4.0 | 4.2 & 0.4 & $H_\mathrm{b1}$ & 10\%            & 8.1  \\ 
4  & 1.7 | 2.0  & 0.3 | 0.7       & 2.7 | 2.3 & 0.4 & $E_\mathrm{b1}$ & 18\%  | 1\%     & 8.4  \\ 
\hline
\end{tabular}
\label{tab:fit_ew}
\end{table*}

\begin{table}[h]
\centering
\caption{Alternative identification model predictions}
\begin{tabular}{c|c|c}
\hline
Mode & $\lambda^*_\textrm{pred}$ & $\lambda^1_{n,m}$ \\
\hline
$H_s$    & $6.7\pm0.7$  & $19.0\pm9.0$ \\ %
$E_s$    & $3.4\pm1.0$  & $6.2\pm3.0$ \\  %
$H_{b1}$ & $2.4\pm0.7$  & $3.7\pm1.8$ \\  %
$E_{b1}$ & $1.9\pm0.6$  & $2.7\pm1.3$ \\  %
\hline
\end{tabular}
\label{tab:pred_ew}
\end{table}

\begin{figure}
\centering
\includegraphics[scale=0.3]{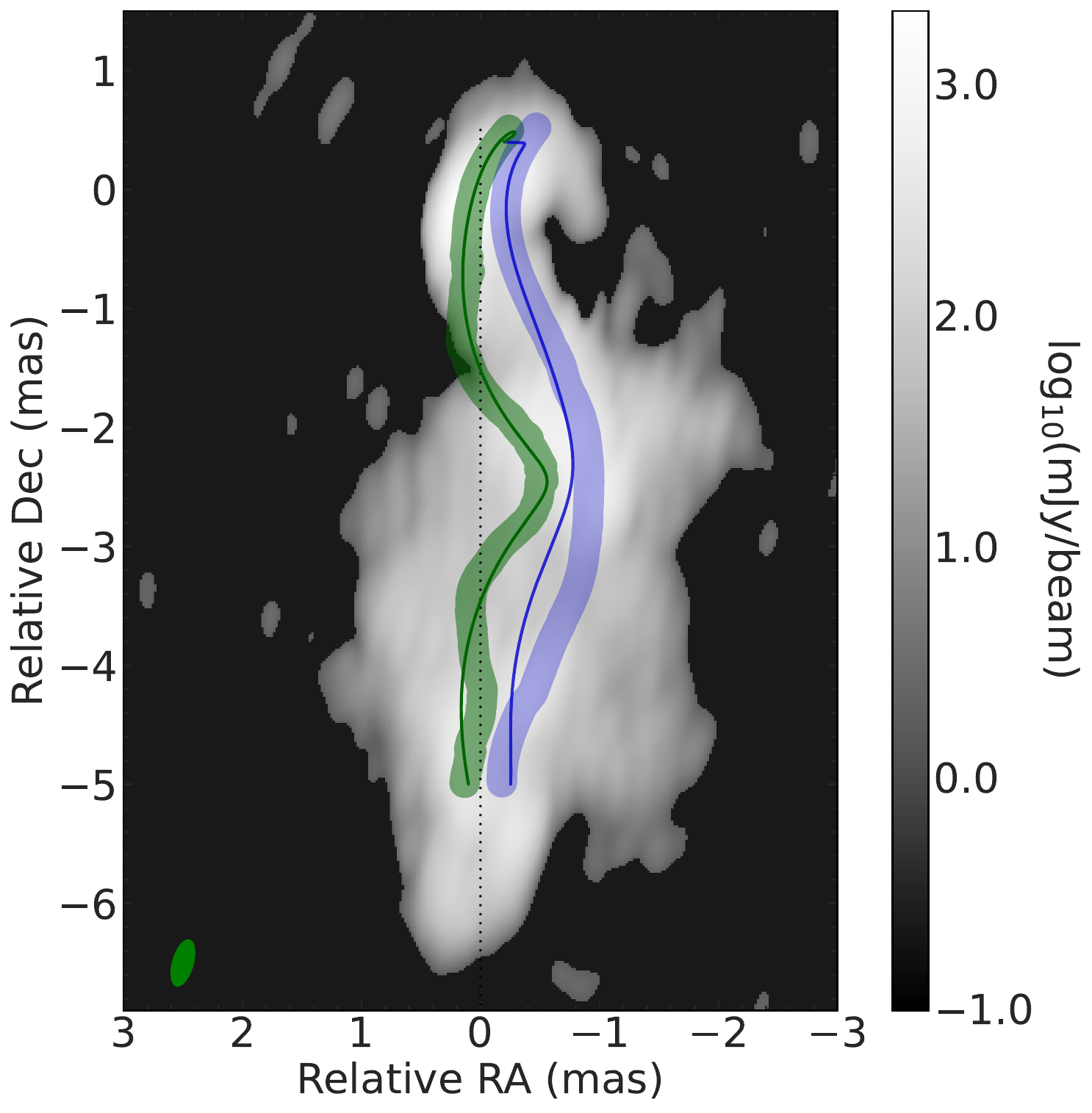}
  \caption{
    Stokes I image of \C\ showcasing an alternative filamentary morphology identification.
    The setup is the same as described in Fig.~\ref{fig:threads}.
    }
     \label{fig:threadsAlt}
\end{figure}

An alternative method of distinguishing between threads is by assigning each fitted Gaussian peak to an `East' and `West' thread, based on their position.
By definition, there is then no overlap between the two filaments and, thus, no helical structure is present, which is the main characteristic of a K-H instability.
The morphology is shown in Fig.~\ref{fig:threadsAlt} and the results of the K-H model fit are displayed in Table~\ref{tab:fit_ew}.
In this case, modelling this morphology with four modes produces a significantly worse fit, as also indicated by the $\chi^2$ values ($\chi_\textrm{amp}^2  \chiamp$ versus $\chi_\textrm{alt}^2  \chialt$).
Thus, higher body modes are required to describe the threads, which are less likely to be present in the flow.
Alternatively, this might be an indication that the K-H description is not well-suited for the thread identification described in this section.
Nevertheless, we note with interest, that the predictions for the alternative model identifications, shown in Table~\ref{tab:pred_ew} are consistent with the ones of the amplitude identification, described in the main text.
This is further indication, that the derived jet parameters describe robustly the outflow in \C.

\end{appendix}

\end{document}